\documentclass{nature}
\usepackage{bm,chngpage,graphicx,mathrsfs}
\makeatletter
\let\saved@includegraphics\includegraphics
\AtBeginDocument{\let\includegraphics\saved@includegraphics}
\renewenvironment*{figure}{\@float{figure}}{\end@float}
\makeatother

\def\apj{Astrophys. J.}
\def\apjl{Astrophys. J. Lett.}
\def\aap{Astron. Astrophys.}
\def\mnras{Mon. Not. Roy. Astron. Soc.}
\def\prl{Phys. Rev. Lett.}
\def\prd{Phys. Rev. D}
\def\nat{Nature}
\def\araa{Ann. Rev. Astron. Astrophys.}

\begin{document}

\title{Repeating Fast Radio Bursts from Collapses of the Crust of a Strange Star}

\author{Jin-Jun Geng$^{1,\ast}$, Bing Li$^{2,3}$, \& Yong-Feng Huang$^{4,5,\ast\ast}$}

\date{\today}{}

\maketitle

\begin{affiliations}
\item Purple Mountain Observatory, Chinese Academy of Sciences, Nanjing 210023, China
\item Key Laboratory of Particle Astrophysics, Institute of High Energy Physics, Chinese Academy of Sciences, Beijing 100049, China
\item Particle Astrophysics Division, Institute of High Energy Physics, Chinese Academy of Sciences, Beijing 100049, China
\item School of Astronomy and Space Science, Nanjing University, Nanjing 210023, China
\item Key Laboratory of Modern Astronomy and Astrophysics (Nanjing University), Ministry of Education, Nanjing 210023, China
\end{affiliations}

\let\thefootnote\relax\footnote{
$^\ast$Correspondence: jjgeng@pmo.ac.cn (J.J.G.); hyf@nju.edu.cn (Y.F.H.)
}

\begin{abstract}

Strange stars (SSs) are compact objects made of deconfined quarks\cite{f1}.
It is hard to distinguish SSs from neutron stars as a thin crust composed of normal hadronic matter
may exist and obscure the whole surface of the SS\cite{f2}.
Here we suggest that the intriguing repeating fast radio bursts (FRBs)
are produced by the intermittent fractional collapses of the crust of an SS
induced by refilling of accretion materials from its low-mass companion.
The periodic/sporadic/clustered temporal behaviors of FRBs could be well understood in our scenario.
Especially, the periodicity is attributed to the modulation of accretion rate through the disk instabilities.
To account for a $\sim 16$-day periodicity of the repeating FRB source 180916.J0158+65,
a Shakura-Sunyaev disk with a viscosity parameter of $\alpha \simeq 0.004$ and an accretion rate
of $\simeq 3 \times 10^{16}$~g~s$^{-1}$ in the low state is invoked.
Our scenario, if favored by future observations, will serve as indirect evidence for the strange quark matter hypothesis.

\end{abstract}

\noindent {\bf INTRODUCTION}

Fast radio bursts (FRBs) are short (duration of $\sim$ ms) radio bursts remaining mysterious.
Its high brightness temperature makes astrophysicists believe that the radio pulse
originates from the so-called coherent emission like the radio pulsar does\cite{f3,f4},
although details for such collective radio emission are still unclear\cite{f5,f6}.
Soon after the first discovery\cite{f7}, rapid progress has been made in the FRB field.
Along with the increasing FRB number, it was found that some of them
are individual events, i.e., do not repeat within a monitor period,
while some sources are repeating since discovery\cite{f8}.

The majority of repeating FRBs have been observed to appear sporadically (e.g., FRB 121102\cite{f9}).
Recently, it is found that the repeating FRB source 180916.J0158+65 (FRB 180916) shows
a $\sim 16$-day periodicity, within which the bursts gather in a 5-day active window\cite{f10}.
Besides, detection of tentative periodic behavior ($\sim 160$ days)
of FRB 121102 over five years of data is reported\cite{f11},
although bursts from FRB 121102 are previously thought to be clustered without a regular pattern.
Several scenarios have been proposed to explain the periodicity,
e.g., the orbital motion of the FRB source\cite{f12,f13,f14,f15,f16}, long-lived precession of the emitting region\cite{f17,f18},
and the ultralong rotational periods of the bursting source\cite{f19}.
However, there is still no consensus on the cause of the periodicity.
An extremely high eccentricity ($e > 0.95$) is required if the
periodicity and the active window come from the pure binary orbital modulation\cite{f14}.
Moreover, it is suggested that theories where the FRB periodicity
arises from forced precession of a magnetar by a companion or fallback disk
are not favored by analyzing the chromatic active window of FRB 180916\cite{f20,f21}.

Putting aside the question that whether repeating and ``non-repeating'' bursts
come from the same physical sources/processes, or are simply due to observational bias,
it is urgent to establish a physical scenario that could naturally account for most characteristics of repeating FRBs.
The coincidence of FRB 200428 and soft gamma-ray repeater 1935+2154 strongly supports that
magnetars are the source of at least some FRBs\cite{f22}.
Motivated by the reasonable explanation for FRB 200428 through the instant accretion onto the magnetar\cite{f23},
we invoke a similar compact object in a binary system to explain the periodic FRB.

\noindent {\bf RESULTS AND DISCUSSION}

\noindent \textbf{Collapse of Crust}

Strange quark matter may be the true ground state of hadronic matter\cite{f1},
from which strange stars (SSs) made of deconfined quarks are predicted.
A thin crust composed of normal hadronic matter may exist and obscure the whole surface of the SS\cite{f2}.
A typical SS with a mass of 1.4 solar mass ($M_{\odot}$)
can not have a crust more massive than $M_{\rm th} \sim 3.4 \times 10^{-6} M_{\odot}$
with a thickness of $\Delta \sim 10^4$~cm and a bottom density of $\rho_{\rm bot} \sim 8.3 \times 10^{10}$~g~cm$^{-3}$~(see Method).

\begin{figure}
	\begin{center}
		\includegraphics[width=\linewidth]{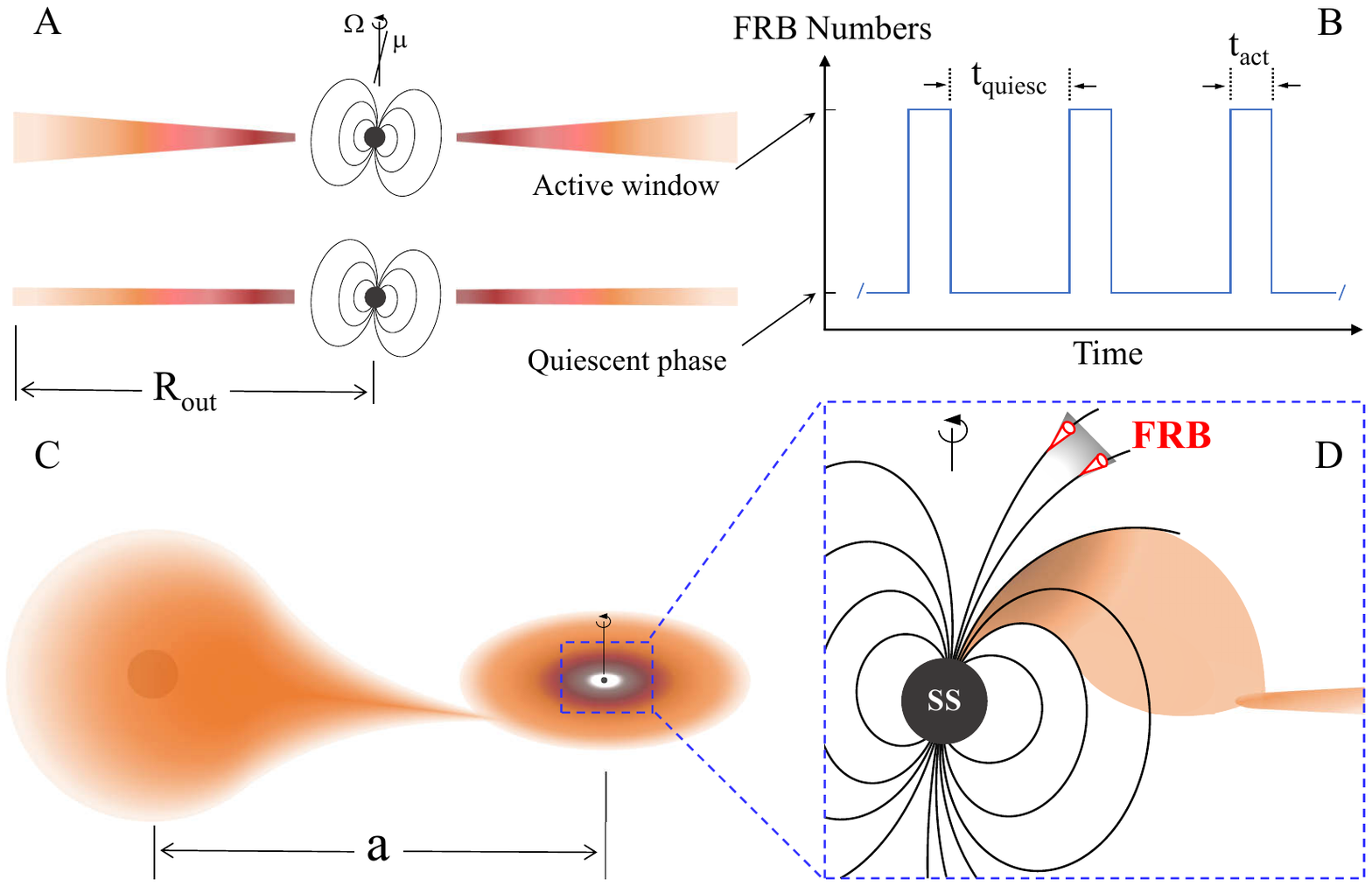}
		\caption{{\bf Schematic illustration of periodic repeating FRBs
		in the SS crust collapse scenario}
        The active window and the quiescent phase of repeating FRBs correspond to
        different states of accreting disk driven by instability (A and B).
        (C and D) The material from the companion is accreted and
        flow along field lines onto the polar cap region of the SS.
        The FRB is produced along open field lines by the plasma cloud launched
        after the fractional collapse of the crust.
        Due to the collisional diffusion at the base of each stream,
        the radius of the collapsed region is larger than the initial radius of the stream (Equation (7)).
        Thus the pairs could flow along the magnetic field lines that are neighbor to
        the field lines guiding the incoming mass stream.}
		\label{fig:Schematic}
	\end{center}
\end{figure}

Like the NS in the low-mass X-ray binary (LMXB), we consider an SS in a binary
with a companion like a K or M dwarf, or a white dwarf.
When materials from the companion's Roche lobe are accreted, a disk around the SS is formed.
As the material gets much closer to the SS, the magnetic field begins to disturb the inflow at
the so-called Alfv\'en radius. Within the Alfv\'en radius, the material is expected to flow
along field lines onto the polar cap region of the SS (Figure~\ref{fig:Schematic}).
Whenever the crust mass exceeds $M_{\rm th}$ or the bottom density exceeds the threshold value, the crust will collapse.
Since the mass inflow may be elongated ``tongues'' like\cite{f24} and
the diffusion timescale is much longer than the free-fall timescale,
only a fraction of the crust near the footprint of these field lines will collapse (See Method).
During each fractional collapse, the sustaining electric field at the crust bottom will be ``turned off''
due to the screen of the electron-positron ($e^+e^-$) pairs from the increasingly heated SS surface (see below),
which will last until the outgoing energy flux is high enough to push away the rest upper crust matter.
Materials from the surrounding disk can continuously refill the crust,
which results in repeating fractional collapses.
We argue that this scenario could naturally explain the repeating FRBs.

Assuming the collapse to be a free-fall process, the collapse timescale is then
$\sqrt{3 \pi / (32 G \rho_{\rm bot})} \sim$ a few ms.
Two kinds of energy, gravitational energy and deconfinement energy are subsequently released after the collapse.
We take the energy rate as $\sim 6.3$ MeV per nucleon\cite{f25,f26} in our following estimates.
For a typical FRB of isotropic energy of $E_{\rm iso} \sim 10^{40}$ erg,
the ratio between the required fallen mass $\delta M$ and the crust mass is
$\delta M/M_{\rm th} \simeq 10^{-10} f_{\mathrm{b},-3} E_{\mathrm{iso},40}$,
where $f_{\rm b}$ is the beaming factor of the FRB.
The convention $Q_x = Q/10^x$ in cgs units is adopted hereafter unless specifically stated.
This means that the crust is an ideal energy reservoir for frequent repeating FRBs.

Usov discussed bare SSs and found that for electromagnetic waves propagating in the hot strange quark matter,
only high-frequency photons ($> 18.5$ MeV) could be efficiently emitted\cite{f27,f28},
i.e., the bare SS surface is a dim X-ray source.
It was thus pointed out that the emission from the heated SS surface should be
dominated by the $e^+e^-$ pairs created in an extremely strong electric field at the surface.
The flux of $e^+e^-$ pairs from unit SS surface is estimated as\cite{f28}
\begin{equation}
\dot{n}_{\pm} \simeq 10^{39} \left( \frac{T_{\rm s}}{10^9~\mathrm{K}} \right)^3 \exp\left[ -11.9 \left(\frac{T_{\mathrm{s}}}{10^9~\mathrm{K}}\right)^{-1} \right]~\mathrm{s}^{-1}~\mathrm{cm}^{-2},
\label{eq:flux}
\end{equation}
where $T_{\rm s}$ is the surface temperature of SS.
Its corresponding energy flux per unit surface is $F_{\pm} \simeq (m_e c^2 + k_{\rm b} T_{\rm s}) \dot{n}_{\pm}$,
where $m_e$ is the electron mass, $c$ is the speed of light, and $k_{\rm b}$ is the Boltzmann constant.
However, due to the extremely high pair density, the pairs will annihilate into photons,
which results in a photon-lepton fireball streaming out along the narrow tunnel of the crust.
The timescale $t_{\rm ann} \sim (\dot{n}_{\pm} \sigma_{\rm T})^{-1}$
for annihilation of $e^+e^-$ pairs is much shorter than the escaping timescale $t_{\rm esc} \sim \Delta / c$,
i.e., $t_{\rm ann} / t_{\rm esc} \le 10^{-4}$ for $T_{\rm s} \ge 10^9$~K, where $\sigma_{\rm T}$ is the Thomson cross section.
Like a bullet in the gun-barrel, $e^+e^-$ pairs will be accelerated by the radiation pressure.
Denoting the number density of $e^+e^-$ pairs as $n_{\pm}$,
the mean free path of these pairs writes as $(n_{\pm} \sigma_{\rm T})^{-1}$.
Considering that the mean free path may be of order $\Delta$ for the pairs to escape, 
we could obtain the density of pairs as $n_{\pm} \simeq (\Delta \sigma_{\rm T})^{-1}$
when they break out the crust.
Assuming $F_{\pm}$ is fully converted into kinetic energy of the pairs, 
the bulk Lorentz factor of the escaping pairs is estimated to be 
$\gamma_{\pm} \simeq \sigma_{\rm T} \Delta F_{\pm} / (m_e c^3)$.
Accurate heat transfer calculation gives that $T_{\rm s}$ is $\sim 10^9$~K
within 10~ms after the crust collapses\cite{f29},
which means that the typical value of $\gamma_{\pm}$ is $\simeq 2000$.
As shown in Method, this relativistic $e^+e^-$ flow could account for the coherent emission of FRBs.

\noindent \textbf{Periodic/Sporadic/Clustered Behavior of FRBs}

The $\sim 16$-day periodicity of FRB 180916 consists of an active duration
of $t_{\rm act} \simeq 5$~days and a quiescence duration of $t_{\rm quiesc} \simeq 11$~days.
In our model, both timescales could be understood by the well-known cycle of the accretion disk
driven by the thermal-viscous instability~(Figure~\ref{fig:Schematic}).
The standard thin disk is thermally (and viscously) unstable for a disk temperature of $\sim 10^4$~K
due to rapid change in opacity associated with the ionization of hydrogen atoms.
A well-defined duty cycle of low/high accretion state is expected from this instability,
and is often invoked for periodic outbursts in LMXB transient systems\cite{f30}.

During the active window, an enhanced accretion rate of $\dot{M}_{\rm act} \simeq 10^{18}$~g~s$^{-1}$ (equivalent to $10^{-8}~M_{\odot}~\mathrm{yr}^{-1}$)
onto the SS is required to repeatedly produce FRBs on the timescale of seconds.
Assuming a Shakura-Sunyaev disk\cite{f31} with a total mass of $M_{\rm disk}$, then the duration of the high accretion state
could be estimated by the depletion timescale of the disk, i.e.,
\begin{eqnarray}
\label{eq:t_act}
t_{\rm act} &=& \frac{M_{\rm disk}}{\dot{M}_{\rm act}} = \frac{2 \pi \int_{R_{\rm in}}^{R_{\rm out}} R \Sigma d R}{\dot{M}_{\rm act}} \\ \nonumber
&\approx& 1.3 \left( \frac{\alpha}{0.01} \right)^{-4/5}
\left( \frac{\dot{M}}{10^{16}~\mathrm{g}~\mathrm{s}^{-1}} \right)^{7/10}
\left( \frac{\dot{M}_{\rm act}}{10^{18}~\mathrm{g}~\mathrm{s}^{-1}} \right)^{-1}
\left( \frac{M_{\mathrm{SS}}}{1.4~M_{\odot}} \right)^{1/4}
\left( \frac{R_{\mathrm{out}}}{10^{10}~\mathrm{cm}} \right)^{5/4}~\mathrm{days},
\end{eqnarray}
where $\alpha$ is the prescription parameter for the viscosity, $\Sigma$ is the surface density,
$\dot{M}$ is the constant disk accretion rate in the low state, $M_{\rm SS}$ is the SS mass,
and $R_{\rm out}$ is the outer edge of the disk, which is much larger than the inner edge $R_{\rm in}$.
In this equation, we have used the dependence of $\Sigma$ (on $\alpha$, $R$, etc) for a standard Shakura-Sunyaev disk\cite{f32}.
A typical value of $0.01$ is used for $\alpha$ here as it corresponds to
the cold phase (low-accretion state) of the disk in thermal-viscous instability theory\cite{f30},
rather than the hot phase during which $\alpha$ is typically assumed to be $0.1$.

\begin{figure}
	\begin{center}
		\includegraphics[width=\textwidth]{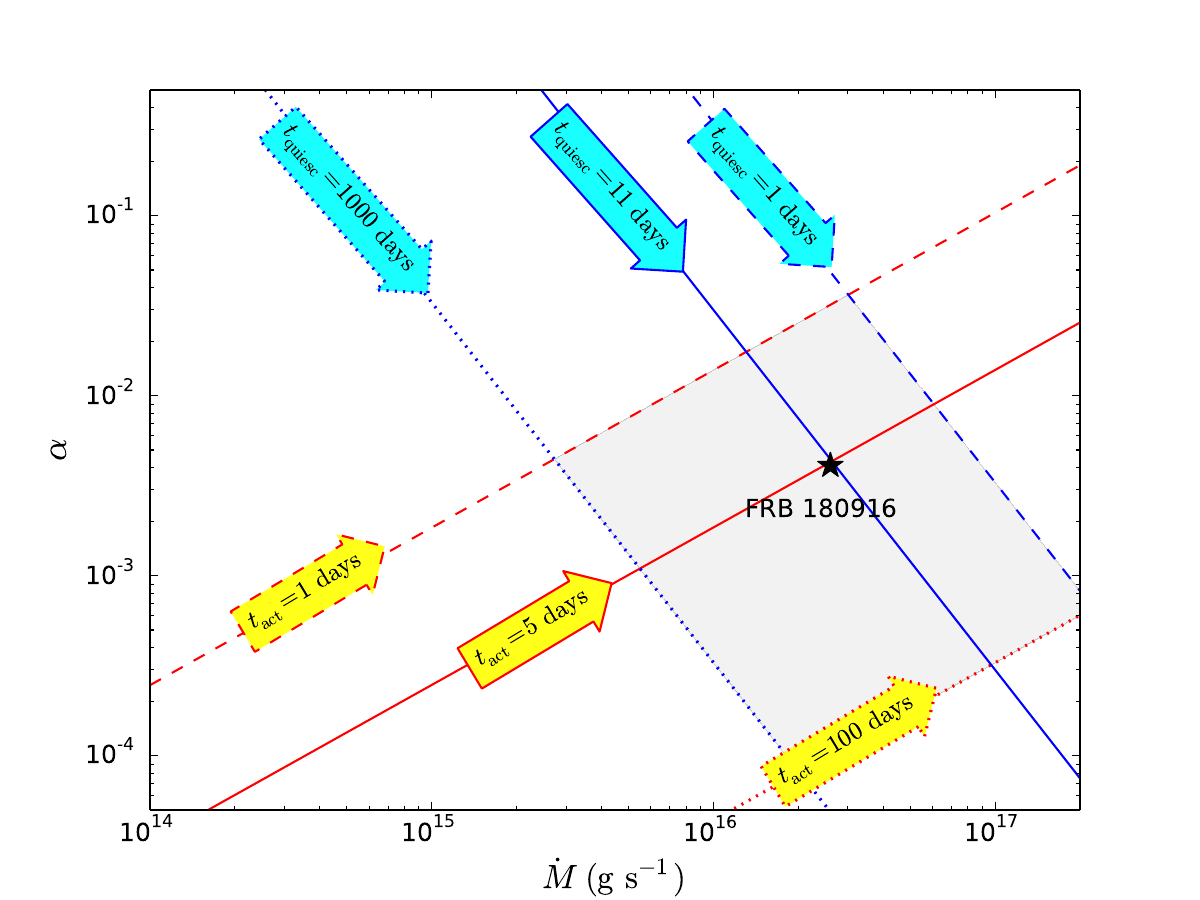}
		\caption{{\bf Constraining disk properties from temporal characteristics of periodic FRBs}
		The red lines present the $\alpha-\dot{M}$ couples for specific $t_{\rm act}$ values derived from Equation (\ref{eq:t_act}),
		while the blues are the corresponding $\alpha-\dot{M}$ couples derived from Equation (\ref{eq:t_quiesc}).
		The shadowed region is the plausible region constrained by taking $t_{\rm act}$ and $t_{\rm quiesc}$
		to be $[1,100]$~days and $[1,1000]$~days respectively.
		The position of the parameters inferred for FRB 180916 is marked by a star symbol.
		}
		\label{fig:Constrain}
	\end{center}
\end{figure}

After the active phase, the disk enters the quiescent phase, during which
matters accumulate up to the critical density so that another instability cycle follows.
A very good estimate of the quiescent duration is given by\cite{f30}
\begin{eqnarray}
\label{eq:t_quiesc}
t_{\rm quiesc} &\approx& 33~\left(\frac{\alpha}{0.01}\right)^{-1} \left(\frac{\dot{M}}{10^{16}~\mathrm{g}~\mathrm{s}^{-1}} \right)^{-2}
\left(\frac{T_c}{3000~\mathrm{K}} \right)^{-1} \\ \nonumber
&& \times \left(\frac{M_{\mathrm{SS}}}{1.4~M_{\odot}} \right)^{-1.26} \left(\frac{R_{\mathrm{out}}}{10^{10}~\mathrm{cm}} \right)^{5.8}~\mathrm{days},
\end{eqnarray}
where $T_c$ is the midplane temperature and is typically suggested to be $\sim 3000$~K at $R_{\rm out} \simeq 10^{10}$~cm.
Using Equations (\ref{eq:t_act}-\ref{eq:t_quiesc}), we could infer that the combination of $\alpha \simeq 0.004$ and $\dot{M} \simeq 3 \times 10^{16}$~g~s$^{-1}$
could match the temporal characteristics of FRB 180916, as shown in Figure~\ref{fig:Constrain}.

The outer edge of the disk can be written as\cite{f33}
\begin{equation}
\label{eq:R_out}
\frac{R_{\rm out}}{a} = (1.0+q)(0.5 - 0.227 \ln q)^4 = F(q),
\end{equation}
where $a = (G M_{\rm SS} / 4 \pi^2)^{1/3} P_{\rm orb}^{2/3}$ is the binary separation
and $q = M_2/M_{\rm SS}$ is the mass ratio of the system.
From Equation (\ref{eq:R_out}) we could further obtain
\begin{equation}
\label{eq:P_orb}
P_{\rm orb} \approx 8~F(q)^{-3/2} \left(\frac{R_{\mathrm{out}}}{10^{10}~\mathrm{cm}}\right)^{3/2}
\left(\frac{M_{\mathrm{SS}}}{1.4~M_{\odot}} \right)^{-1/2}~\mathrm{min}.
\end{equation}
Taking $q = 0.4$ in the binary system for FRB 180916,
the orbital period of this system could be inferred as 40~min.
Such a binary system is like some ultracompact X-ray sources\cite{f34,f35},
but the primary star here is an SS rather than an NS.
Since the FRB pulses are observed along the open magnetic field lines,
the binary here is preferred to be face-on rather than edge-on. 
As a result, FRB signals would not be markedly modulated by the orbital motion 
since no eclipse is expected in a face-on binary system.
However, orbital period modulation might still be inferred when such an
FRB source is in our galaxy or a very nearby galaxy
so that an X-ray or even optical counterpart could be detected. 
The donor could be either a white dwarf or a low-mass star
since the general trend in the stability of hydrogen-poor accretion disks is similar to that of a hydrogen disk\cite{f35}.
If we assume that different-frequency emissions of FRBs are produced at different altitudes within the pulsar magnetosphere\cite{f21},
the observed chromatic activity window could be explained by the drift of the main collapse regions during the high accretion state.
Such a drift is expected considering that it takes some extra waiting time for regions that have just collapsed to refill the materials
to reach the critical state. Moreover, there may be a drift for the landing points of the accretion streams
due to the evolution of the Alfv\'en radius induced by the variation of $\dot{M}_{\rm act}$ during $t_{\rm act}$.

Except for the periodic behavior, some repeating FRBs have shown the sporadic/clustered behavior.
These characteristics could also be naturally interpreted in our scenario.
For a relatively long $P_{\rm orb}$,
the accretion rate during the quiescence is expected to become relatively small,
the quiescent duration will be increasingly longer due to its dependence on $\dot{M}$ in Equation (\ref{eq:t_quiesc}).
On the other hand, the active window will be decreased,
which could result in clustered bursts in long quiescence.
In an even smaller $\dot{M}$ system, it takes a longer time to refill the crust of the SS
to reach the critical point. At the same time, the smaller total disk mass to be accreted onto
the crust may supply only one or several radio bursts in each active window.
This situation should correspond to the case of the sporadic repeating FRBs.

In our scenario, FRB 180916 is supposed to be in a binary system with a very short
orbital period, which might be rare even among LMXBs.
Therefore, periodic repeating FRBs may not be common in repeating FRBs.
Lessons from studies on outbursts of LMXB transients
indicate that mass-transfer rate variations, disk heating, irradiation,
and other physical processes can produce the wealth of low-mass X-ray binary outburst light-curves\cite{f30}.
Long-term monitoring on a large sample of FRBs will help to reveal this temporal similarity to LMXBs.

\noindent \textbf{Conclusion}

We propose that repeating FRBs may be generated from the fractional collapses of the crust of the SS.
In our scenario, each FRB pulse is released along the open field lines
above the polar cap of the SS after the crust matter collapse.
The refilling of the accretion materials will repeat this process and hence lead to repeating FRBs.
The periodic/sporadic/clustered behaviors of repeating FRBs
are naturally explained by the accretion modulation in the binary system.
Moreover, we could infer the basic parameters of the disk from the active/quiescent duration of the bursts.

Comparing with the NS accretion process, accretion onto an SS with a crust makes the repeats feasible. 
An SS may form if the gravity becomes strong enough to deconfine neutron matter into
strange quark matter (without going too far and collapsing to a black hole),
which could happen within a star through an extremely energetic supernova (e.g., a hypernova),
an accreting NS in X-ray binaries, or mergers of two NSs.
The exact transition point between degenerate neutron matter and quark matter is unknown
because of the lack of a clear understanding of the strong force and quark matter.
We suggest that the SS in the binary here may come from the core-collapse supernova  
of a progenitor star with a relatively larger mass in the range that is thought to give birth to an NS (8-30 $M_{\odot}$).
The accurate ratio of SSs to NSs in the Universe is unknown,
so our scenario predicts that repeating FRBs may be associated with only some (NOT all) face-on LMXBs.
The strange quark matter hypothesis dates back to the early 1970s and has neither been proven nor disproven.
If our scenario is favored in the future, it could serve as indirect evidence for the existence of SS
and support for the strange quark matter hypothesis.
Thus we strongly encourage monitoring the face-on LMXBs for future repeating FRB searches.

\noindent {\bf METHOD}

\noindent \textbf{Strange Star with Crust}

At the bare SS surface, the quarks bound by the confinement of strong interaction have
a very sharp surface with a thickness of the order of 1 fermi, i.e.,
the density changes abruptly from $\sim 5 \times 10^{14}$~g~cm$^{-3}$ to zero.
On the contrary, the electrons bound by the Coulomb force can extend several hundred fermis beyond the quark surface.
As a consequence, in a thin layer of several hundred fermis thick above the strange matter surface,
a strong electric field is established, which is estimated to be about $10^{17}$~V~cm$^{-1}$ and outwardly directed.
This large outward-directed electric field will support some accreted normal materials,
which results in a thin crust covering the SS\cite{f36}.

Huang \& Lu calculated the balance between the electrical and gravitational forces across the thin gap
and find that the maximum density at the base of the crust is only one-fifth of the neutron drip point\cite{f37}.
So the bottom material of the crust can not convert into free neutrons and gradually flow onto the SS.
Otherwise, when the crust gets heavier and heavier, the gap will decrease to an extent so that the crust collapse occurs.
Accurate calculations give that the maximum crust mass is $M_{\rm th} \sim 3.4 \times 10^{-6} M_{\odot}$
and its corresponding thickness is $10^4$~cm\cite{f37}.

\noindent \textbf{Accretion and Crust Collapse}

It has been proposed that a crust may not form after the birth of an SS if the
SS is rapidly rotating\cite{f38,f39}.
However, the formation of the crust could be guaranteed by
the accretion from the companion, as occurred in our framework.
For a dipole magnetic field with a surface strength of $B_{\rm SS}$,
the condition for magnetically-funneled column accretion onto the SS is that
the Alfv\'en radius is less than the co-rotating radius, which requires
\begin{equation}
B_{\rm SS} < 1.0 \times 10^{13} M_{\rm{SS}}^{5/6} \dot{M}_{18}^{1/2} R_{\mathrm{SS},6}^{-3} P_{\mathrm{SS},0}^{7/6}~\mathrm{G},
\end{equation}
where $R_{\rm SS}$ and $P_{\rm SS}$ are the SS radius and spin period respectively.
This requirement could be met even for ``SS magnetars'' if the SS is slowly rotating ($P_{\rm SS} \ge 10$~s) and the accretion rate is high.
Moreover, the surface magnetic field near the pole region may be dominated by a multipolar component
that could be about an order of magnitude larger than the dipole field component,
which has been inferred in some pulsating ultraluminous X-ray sources\cite{f40}.
Therefore, the crust of the SS could be formed from materials accreted from the companion.
On the other hand, when the accretion rate is low (e.g., $\sim 10^{16}$~g~s$^{-1}$),
the Alfv\'en radius will increase and the accretion system may enter the propeller regime,
which corresponds to the quiescence phase of repeating FRBs since no materials will be accreted onto the SS.

While the accreted material is thought to flow towards the polar cap region, we lack the knowledge of its geometry.
Here, we assume the flow should be elongated ``tongues'' like and hit the SS in random places\cite{f24}.
Taking $\zeta$ as the filling factor of the column and the number of simultaneous streams to be $N$,
the sectional area of each stream is $\zeta A / N$, where $A$ is the polar cap area.
So materials accumulate significantly at some points of the crust surface.
These regions will collapse when the local pressure at the base of the crust exceeds the critical stress.
At the base of the stream, the transverse velocity resulted from collisional diffusion in the presence of a pressure
gradient ($\nabla P$) could be approximated by\cite{f25,f41}
$v_{\rm d} \sim 1.3 \times 10^{-24} T_{6}^{-3/2} B_{\rm{SS},14}^{-2} \nabla P~\mathrm{cm}~\mathrm{s}^{-1}$ ($\nabla P$ in unit of erg~cm$^{-4}$),
where $T$ is the temperature of the surface matter, $\nabla P \sim \rho g h / r_{\rm d}$,
$g = G M_{\rm SS} / R_{\rm SS}$ is the surface gravity, $h$ is the column height,
and $r_{\rm d}$ is the spreading radius of the stream.
The density of each stream near the crust surface reads $\rho = \dot{M} / (\zeta A v_{\rm fd})$,
where $v_{\rm fd} = (2 G M_{\rm SS} / R_{\rm SS})^{1/2}/2$ is the velocity of matter that falls down.
During the collapse timescale of $\delta t$,
the spreading radius of the stream could be calculated as $r_{\rm d} = v_{\rm d} \delta t$.
Using the closure relation of $N \pi r_{\rm d}^2 \simeq \zeta A$, we obtain the expression of $\zeta$
and hence the mass of the crust that would collapse to be
\begin{equation}
\delta m = \frac{\pi r_{\rm d}^2}{4 \pi R_{\rm SS}^2} M_{\rm th}
\approx 1.0 \times 10^{18} N^{-1/2} T_{6}^{-3/4} B_{\mathrm{SS,14}}^{-1} \dot{M}_{18}^{1/2} h_4^{1/2} \delta t_{5 \mathrm{ms}}^{1/2}~\mathrm{g}.
\end{equation}
Therefore, the amount of mass for each fractional collapse is $\le 10^{18}$~g for $N > 1$.

It should be noted that there may be several regional collapses within a short
waiting time as $N > 1$ and each collapse is independent in principle, i.e.,
the waiting time between the pulses of repeating FRBs could be within only several seconds or shorter.
However, this situation should not be frequent. If the instant collapse rate is high so that
the crust of the SS loses a significant fraction of its mass in a short duration, 
the created tunnels would absorb materials from neighboring regions. 
Then to produce the next burst cluster, it will take a longer time to accrete enough mass to
reach the critical state, with the equivalent isotropic crust mass being no less than $M_{\rm th}$.
On average, an accretion rate of $\sim 10^{18}$~g~s$^{-1}$ is required to repeatedly produce FRBs on the timescale of seconds.

\noindent \textbf{Coherent Emission}

Both pulsar-like mechanisms that invoke the magnetosphere of a compact object\cite{f4,f42,f43} and
the synchrotron maser mechanism in relativistic shocks\cite{f44} could in principle (still in debate) produce coherent radio bursts.
Currently, the polarization analyses on repeating FRB 180301\cite{f45} and
the chromatic activity window of FRB 180916 support a magnetospheric origin for the radio emission.
After the crust collapse, two possible ways, i.e., the pair spraying from the tunnel
and the Alfv\'en wave propagating in the charge-deficit region\cite{f46} could produce pair bunching.
It is difficult to clarify which mechanism works for FRBs in the real situation in this report.
Here, we adopt the prior one in our scenario and show its self-consistence.
As shown in Figure~\ref{fig:Schematic}, the polar cap region of the crust will collapse due to the
continuous mass increasing from accretion, which results in a relativistic $e^+e^-$ flow
streaming outwards along the open field lines.
There are relatively cleaner spaces in the vicinities of the ``tongues'' like streams.
Even when the accretion flow is assumed to be homogeneous in the environment above the polar cap,
the ratio of the outgoing released energy flux $F_{\pm}$ to the kinetic energy flux of the accretion flow ($\dot{M} v_{\rm fd}^2 / A$)
is greater than 1 for $T_{\rm s} > 1.3 \times 10^9$~K according to Equation (1).
Thus the $e^+e^-$ flow could penetrate through the accretion flow.
The parallel electric field above the crust is screened during the pair spraying.
When these pairs leave the magnetosphere gap region, the gap electric field will 
initiate again within a short period of $\sim \mu$s,
which produces another following plasma cloud with a bulk Lorentz factor $>10^4$ by pulsar-like sparking\cite{f3}.
The following plasma cloud will catch up with the plasma cloud emitted from collapse,
and lead to the two-stream instability in the plasma, hence the coherent emission.
The characteristics of the two plasma clouds are discussed below.

Let's consider a sparking gap with a thickness of $H$ above the polar cap of the SS magnetosphere.
The condition that curvature radiation photons produce a pair in the gap is\cite{f3} \footnote{
In the original formulation\cite{f3}, the term of $B_{\perp}/B_q$ is used
rather than $B_{\rm SS}/B_q$ in the left side of the equality,
where $B_{\perp} \sim B_{\rm SS} H / R_{\rm SS}$ is the perpendicular (to the direction of propagation for the photon) component of the surface magnetic field.
However, in relevant calculations and later literature,
it has been widely assumed that the typical radius of curvature of the magnetic field lines very near the pulsar surface
is roughly $R_{\rm SS}$, i.e., higher multipole components contribute most strongly near the surface rather than dipolar.
Thus in order to maintain the self-consistency, it is reasonable to take $B_{\perp} \simeq B_{\rm SS}$.}
\begin{equation}
\frac{3 \hbar}{4 \pi m_e c R_{\rm SS}} \left( \frac{q_e \Omega B_{\rm SS} H^2}{m_e c^3} \right)^3 \frac{B_{\rm SS}}{B_q} \approx \frac{1}{15},
\end{equation}
where $\hbar$ is the reduced Planck's constant, $q_e$ is the electron charge,
$\Omega = 2 \pi / P_{\rm SS}$ is the SS angular velocity,
and $B_q = m_e^2 c^3 / (q_e \hbar)$. The gap height is then derived as
\begin{equation}
H \approx 100~R_{\rm{SS},6}^{1/6} P_{\rm{SS},0}^{1/2} B_{\rm{SS},14}^{-2/3}~\mathrm{cm}.
\label{eq:H}
\end{equation}

Assuming the bulk Lorentz factor of the prior plasma cloud is $\gamma$,
then the overlapping radius\cite{f23,f47} for the two clouds is $r_{\rm emi} \simeq 2 \gamma^2 H$.
In a dipolar field, the curvature radius $r_{\rm c}$ is related to $r_{\rm emi}$ by $r_{\rm c} \simeq 4 r_{\rm emi}/(3 \sin \theta_{\rm emi})$,
where $\theta_{\rm emi}$ is the poloidal angle of the emission region.
The corresponding characteristic frequency of curvature emission is
\begin{equation}
\nu_{\rm c} = \gamma^3 \frac{3 c}{4 \pi r_{\rm c}}.
\end{equation}
Therefore, the Lorentz factor of the prior plasma cloud needs to be
\begin{equation}
\gamma = 1040 \left(\frac{\nu_{\rm c}}{1.4~\mathrm{GHz}}\right) \left(\frac{H}{100~\mathrm{cm}}\right) \left(\frac{\sin \theta_{\mathrm{emi}}}{0.05}\right)^{-1}.
\label{eq:gamma}
\end{equation}
This value is generally consistent with that of pairs leaving from the SS mentioned above.
Equations (\ref{eq:H}) and (\ref{eq:gamma}) indicate that a relatively high local magnetic field of $> 10^{13}$
is preferred in the bunching scenario to make sure that $r_{\rm emi}$ is within the light cylinder.

According to relevant calculations of the coherent emission\cite{f23},
the electron number density of the emitting region is
\begin{eqnarray}
n_e &\simeq& 1.4 \times 10^{10}
\left( \frac{L_{\rm FRB}}{1 \times 10^{43}~\mathrm{erg~s}^{-1}} \right)^{1/2}
\left(\frac{\delta t}{5~\mathrm{ms}} \right)^{-1/2} \\ \nonumber
& & \times \left(\frac{\nu_{\rm c}}{1.4 \times 10^9~\mathrm{Hz}} \right)^{-5/2}
\left(\frac{H}{100~\mathrm{cm}} \right)^{-4} \left(\frac{\sin \theta_{\mathrm{emi}}}{0.05} \right)^{2}
\mathrm{cm}^{-3},
\end{eqnarray}
where $L_{\rm FRB}$ is isotropic FRB luminosity, $\delta t$ is the duration of the FRB.
The corresponding plasma density near the SS should be $\sim n_e (R_{\rm SS} / r_{\rm emi})^{-3}$,
which is found to be less than $n_{\pm}$ produced during the collapse for a reasonable range of $H$ and $\theta_{\mathrm{emi}}$.
This indicates that the crust collapse could naturally provide the prior plasma cloud for coherent emission.

\noindent {\bf Acknowledgements}

The authors thank the anonymous referees for their constructive suggestions. 
We also would like to thank Xue-Feng Wu for stimulating discussions.
This work is partially supported
by National SKA Program of China No. 2020SKA0120300,
by the National Natural Science Foundation of China
(grants No. 11903019, 11873030, 11833003, 12041306, U1938201, U1838113),
by the Strategic Priority Research Program of the Chinese Academy of Sciences
(``multi-waveband Gravitational Wave Universe'' grant No. XDB23040000),
and by the science research grants from the China Manned Space Project with NO. CMS-CSST-2021-B11. 

\noindent {\bf Author Contributions}

J.J.G. and Y.F.H. led the project and wrote the manuscript.
J.J.G. performed all the calculations, generated Figure 2, and provided the explanation for periodicity of FRBs.
Y.F.H. suggested the basic physical origin of the FRB emission. B.L. generated Figure 1.
All authors discussed the results and commented on the manuscript.

\noindent {\bf Declaration of Interests}

The authors declare no competing interests.

\noindent {\bf Lead Contact Website}

Jin-Jun Geng: https://orcid.org/0000-0001-9648-7295; Yong-Feng Huang: https://orcid.org/0000-0001-7199-2906.

\clearpage

\clearpage


\noindent {\bf References}
\begin{thebibliography}{10}

\bibitem{f1} Witten, E. (1984). Cosmic separation of phases. \prd~{\it \bf 30}, 272-285, 10.1103/PhysRevD.30.272.

\bibitem{f2} Alcock, C., Farhi, E., and Olinto, A. (1986). Strange Stars. \apj~{\it \bf 310}, 261, 10.1086/164679.

\bibitem{f3} Ruderman, M.A., and Sutherland, P.G. (1975). Theory of pulsars: polar gaps, sparks, and coherent microwave radiation. \apj~{\it \bf 196}, 51-72, 10.1086/153393.

\bibitem{f4} Kumar, P., Lu, W., and Bhattacharya, M. (2017). Fast radio burst source properties and curvature radiation model. \mnras~{\it \bf 468}, 2726-2739, 10.1093/mnras/stx665.

\bibitem{f5} Zhang, B. (2020). The physical mechanisms of fast radio bursts. \nat~{\it \bf 587}, 45-53, 10.1038/s41586-020-2828-1.

\bibitem{f6} Melrose, D.B., Rafat, M.Z., and Mastrano, A. (2021). Pulsar radio emission mechanisms: a critique. \mnras~{\it \bf 500}, 4530-4548, 10.1093/mnras/staa3324.

\bibitem{f7} Lorimer, D.R., Bailes, M., McLaughlin, M.A., et al. (2007). A Bright Millisecond Radio Burst of Extragalactic Origin. Science~{\it \bf 318}, 777, 10.1126/science.1147532.

\bibitem{f8} Cordes, J.M., and Chatterjee, S. (2019). Fast Radio Bursts: An Extragalactic Enigma. \araa~{\it \bf 57}, 417-465, 10.1146/annurev-astro-091918-104501.

\bibitem{f9} Spitler, L.G., Scholz, P., Hessels, J.W.T., et al. (2016). A repeating fast radio burst. \nat~{\it \bf 531}, 202-205, 10.1038/nature17168.

\bibitem{f10} CHIME/FRB Collaboration, Amiri, M., Andersen, B.C., Bandura, et al. (2020). Periodic activity from a fast radio burst source. \nat~{\it \bf 582}, 351-355, 10.1038/s41586-020-2398-2.

\bibitem{f11} Rajwade, K.M., Mickaliger, M.B., Stappers, B.W., et al. (2020). Possible periodic activity in the repeating FRB 121102. \mnras~{\it \bf 495}, 3551-3558, 10.1093/mnras/staa1237.

\bibitem{f12} Dai, Z.G., Wang, J.S., Wu, X.F., and Huang, Y.F. (2016). Repeating Fast Radio Bursts from Highly Magnetized Pulsars Traveling through Asteroid Belts. \apj~{\it \bf 829}, 27, 10.3847/0004-637X/829/1/27.

\bibitem{f13} Zhang, B. (2017). A {\textquotedblleft}Cosmic Comb{\textquotedblright} Model of Fast Radio Bursts. \apjl~{\it \bf 836}, L32, 10.3847/2041-8213/aa5ded.

\bibitem{f14} Gu, W.M., Yi, T., and Liu, T. (2020). A neutron star-white dwarf binary model for periodically active fast radio burst sources. \mnras~{\it \bf 497}, 1543-1546, 10.1093/mnras/staa1914.

\bibitem{f15} Dai, Z.G., and Zhong, S.Q. (2020). Periodic Fast Radio Bursts as a Probe of Extragalactic Asteroid Belts. \apjl~{\it \bf 895}, L1, 10.3847/2041-8213/ab8f2d.

\bibitem{f16} Ioka, K., and Zhang, B. (2020). A Binary Comb Model for Periodic Fast Radio Bursts. \apjl~{\it \bf 893}, L26, 10.3847/2041-8213/ab83fb.

\bibitem{f17} Zanazzi, J.J., and Lai, D. (2020). Periodic Fast Radio Bursts with Neutron Star Free Precession. \apjl~{\it \bf 892}, L15, 10.3847/2041-8213/ab7cdd.

\bibitem{f18} Yang, H., and Zou, Y.C. (2020). Orbit-induced Spin Precession as a Possible Origin for Periodicity in Periodically Repeating Fast Radio Bursts. \apjl~{\it \bf 893}, L31, 10.3847/2041-8213/ab800f.

\bibitem{f19} Beniamini, P., Wadiasingh, Z., and Metzger, B.D. (2020). Periodicity in recurrent fast radio bursts and the origin of ultralong period magnetars. \mnras~{\it \bf 496}, 3390-3401, 10.1093/mnras/staa1783.

\bibitem{f20} Pleunis, Z., Michilli, D., Bassa, C.G., et al. (2021). LOFAR Detection of 110-188 MHz Emission and Frequency-dependent Activity from FRB 20180916B. \apjl~{\it \bf 911}, L3, 10.3847/2041-8213/abec72.   

\bibitem{f21} Li, D.Z., and Zanazzi, J.J. (2021). Emission Properties of Periodic Fast Radio Bursts from the Motion of Magnetars: Testing Dynamical Models. \apjl~{\it \bf 909}, L25, 10.3847/2041-8213/abeaa4.

\bibitem{f22} CHIME/FRB Collaboration, Andersen, B.C., Bandura, K.M., Bhardwaj, M., et al. (2020). A bright millisecond-duration radio burst from a Galactic magnetar. \nat~{\it \bf 587}, 54-58, 10.1038/s41586-020-2863-y.

\bibitem{f23} Geng, J.J., Li, B., Li, L.B., et al. (2020). FRB 200428: An Impact between an Asteroid and a Magnetar. \apjl~{\it \bf 898}, L55, 10.3847/2041-8213/aba83c.

\bibitem{f24} Romanova, M.M., Kulkarni, A.K., and Lovelace, R.V.E. (2008). Unstable Disk Accretion onto Magnetized Stars: First Global Three-dimensional Magnetohydrodynamic Simulations. \apjl~{\it \bf 673}, L171, 10.1086/527298.

\bibitem{f25} Cheng, K.S., Dai, Z.G., Wei, D.M., and Lu, T. (1998). Is GRO J1744-28 a Strange Star? Science~{\it \bf 280}, 407, 10.1126/science.280.5362.407.

\bibitem{f26} Jia, J.J., and Huang, Y.F. (2004). A numerical study of the collapse of the crust of strange stars. Chin. Astron. \& Astroph.~{\it \bf 28}, 144-153, 10.1016/S0275-1062(04)90017-3.

\bibitem{f27} Usov, V.V. (1998). Bare Quark Matter Surfaces of Strange Stars and e$^{+}$e$^{-}$ Emission. \prl~{\it \bf 80}, 230-233, 10.1103/PhysRevLett.80.230.

\bibitem{f28} Usov, V.V. (2001). Thermal Emission from Bare Quark Matter Surfaces of Hot Strange Stars. \apjl~{\it \bf 550}, L179-L182, 10.1086/319639.

\bibitem{f29} Zhang, Y., Geng, J.J., and Huang, Y.F. (2018). Fast Radio Bursts from the Collapse of Strange Star Crusts. \apj~{\it \bf 858}, 88, 10.3847/1538-4357/aabaee.

\bibitem{f30} Lasota, J.P. (2001). The disc instability model of dwarf novae and low-mass X-ray binary transients. New Astron. Rev.~{\it \bf 45}, 449-508, 10.1016/S1387-6473(01)00112-9.

\bibitem{f31} Shakura, N.I., and Sunyaev, R.A. (1973). Black holes in binary systems. Observational appearance. \aap~{\it \bf 500}, 33-51.

\bibitem{f32} Frank, J., King, A., and Raine, D.J. (2002). Accretion Power in Astrophysics: Third Edition (Cambridge University Press).

\bibitem{f33} Paczy{\'n}ski, B. (1977). A model of accretion disks in close binaries. \apj~{\it \bf 216}, 822-826, 10.1086/155526.

\bibitem{f34} Juett, A.M., Psaltis, D., and Chakrabarty, D. (2001). Ultracompact X-Ray Binaries with Neon-rich Degenerate Donors. \apjl~{\it \bf 560}, L59-L63, 10.1086/324225.

\bibitem{f35} Lasota, J.P., Dubus, G., and Kruk, K. (2008). Stability of helium accretion discs in ultracompact binaries. \aap~{\it \bf 486}, 523-528, 10.1051/0004-6361:200809658.

\bibitem{f36} Cheng, K.S., Dai, Z.G., and Lu, T. (1998). Strange Stars and Related Astrophysical Phenomena. Int. J. Mod. Phys. D.~{\it \bf 7}, 139-176, 10.1142/S0218271898000139.

\bibitem{f37} Huang, Y.F., and Lu, T. (1997). Strange stars: how dense can their crust be? \aap~{\it \bf 325}, 189-194.

\bibitem{f38} Xu, R.X., Qiao, G.J., and Zhang, B. (1999). PSR 0943+10: A Bare Strange Star? \apjl~{\it \bf 522}, L109-L112, 10.1086/312226.

\bibitem{f39} Xu, R.X., Zhang, B., and Qiao, G.J. (2001). What if pulsars are born as strange stars? Astropart. Phys.~{\it \bf 15}, 101-120, 10.1016/S0927-6505(00)00154-7.

\bibitem{f40} Israel, G.L., Belfiore, A., and Stella, L., et al. (2017). An accreting pulsar with extreme properties drives an ultraluminous x-ray source in NGC 5907. Science~{\it \bf 355}, 817-819, 10.1126/science.aai8635.

\bibitem{f41} Spitzer, L. (1965). Physics of Fully Ionized Gases (New York: Interscience).

\bibitem{f42} Kumar, P., and Bo{\v{s}}njak, {\v{Z}}. (2020). FRB coherent emission from decay of Alfv{\'e}n waves. \mnras~{\it \bf 494}, 2385-2395, 10.1093/mnras/staa774.

\bibitem{f43} Yang, Y.P., and Zhang, B. (2018). Bunching Coherent Curvature Radiation in Three-dimensional Magnetic Field Geometry: Application to Pulsars and Fast Radio Bursts. \apj~{\it \bf 868}, 31, 10.3847/1538-4357/aae685.

\bibitem{f44} Metzger, B.D., Margalit, B., and Sironi, L. (2019). Fast radio bursts as synchrotron maser emission from decelerating relativistic blast waves. \mnras~{\it \bf 485}, 4091-4106, 10.1093/mnras/stz700.

\bibitem{f45} Luo, R., Wang, B.J., Men, Y.P., et al. (2020). Diverse polarization angle swings from a repeating fast radio burst source. \nat~{\it \bf 586}, 693-696, 10.1038/s41586-020-2827-2.

\bibitem{f46} Lu, W.B., Kumar, P., and Zhang, B. (2020). A unified picture of Galactic and cosmological fast radio bursts. \mnras~{\it \bf 498}, 1397-1405, 10.1093/mnras/staa2450.

\bibitem{f47} Mitra, D. (2017). Nature of Coherent Radio Emission from Pulsars. Journal of Astrophysics and Astronomy~{\it \bf 38}, 52, 10.1007/s12036-017-9457-6.

\end{thebibliography}
\end{document}